\documentclass[12pt]{article}
\begin{document}
\input{epsf}
\Large
\begin{center}
\bf{
    $A$-dependence of coherent electroproduction of $\rho^{0}$ mesons
  on nuclei in forward direction}
\end{center}

\normalsize
\begin{center}
      N.~Akopov,  Z.~Akopov,  G.~Aslanyan$^1$,  L.~Grigoryan\footnote{Supported by DESY, Deutsches
      Elektronen Synchrotron}
\end{center}

\begin{center}
      Yerevan Physics Institute, Br.Alikhanian 2, 375036 Yerevan, Armenia
\end{center}
\begin {abstract}
\hspace*{1em} 
This article presents the $A$-dependence of the differential cross section for the
coherent electroproduction of vector mesons on nuclei in forward direction,
 at fixed values of longitudinal momentum transfer $q_{L}$.
It is shown that such cross section has complicated behavior over the atomic mass number 
$A$ with local minimums and maximums.
It is also shown that a ratio of the real to the imaginary parts of the forward coherent amplitude 
on nuclei $\alpha_{A} = \Re e{f_{A}} / \Im m{f_{A}}$ has
breaking points at some values of $A$.
Comparison of the behaviors of the normalized cross section
$\Big(\frac{d\sigma}{d\Omega}\Big)_{A}\Big/\Big(\frac{d\sigma}{d\Omega}\Big)_{N}$
 and $\alpha_{A}$ over $A$
shows that the location of minimums of the cross section are very close   
to the breaking points of $\alpha_{A}$.
. 
\end {abstract} 
\normalsize
\hspace*{1em} 
Normally the behavior of the cross section for the reactions
on nuclear targets can be presented in the form $\sigma \sim A^{\alpha}$, where $A$ is the
atomic mass number of the nuclear target.
 For many reactions $\alpha$ is
a constant or a weakly changing function of $A$. 
The experimental data on coherent
photoproduction of $\rho^{0}$ mesons on nuclear targets in forward direction, 
in the region of moderate energies of photons~\cite{A1}, indicates that the ratio
$\Big(\frac{1}{A}\frac{d\sigma}{d\Omega}\Big)_{A}\Big/
\Big(\frac{1}{9}\frac{d\sigma}{d\Omega}\Big)_{Be}$ as a function of $A$
increases in the domain of light nuclei, achieves maximum on
medium nuclei $(A \sim 40 \div 60)$ and then decreases. The position of the maximum
depends from the value of the photon energy $\nu$.
When the energy of the photon increases, the position of the maximum shifts to the larger
values of $A$.
This behavior is governed mainly by the longitudinal momentum transfer, $q_{L}$.
For the photoproduction process
$q_{L} = M^{2}_{V}/2\nu$, where $M_{V}$ is the vector meson mass. 
The value of $q_{L}$ quickly decreases with the increase of the photon energy.
In the case of vector mesons electroproduction $q_{L}$ depends from 
$\nu$ and the photon virtuality $Q^{2}$. Contrary to the real
photoproduction case, for values of $Q^{2} > M^{2}_{V}$, $q_{L}$ may differ
from zero in a wide enough range of $\nu$.\\
Calculations performed for the case of coherent electroproduction of
vector mesons on nuclei indicate that the $A$-dependence of 
the coherent differential cross sections have complicated behavior with
minimums and maximums.\\
\hspace*{1em}
We study also a ratio of the real to the imaginary parts of forward amplitude on
nucleus $\alpha_{A} = \Re e{f_{A}} / \Im m{f_{A}}$ as a function of $A$.
It is shown that $\alpha_{A}$ has 
breaking points, i.e. jumps from $- \infty$
to $\infty$ at some values of $A$. It happens when imaginary part of nuclear
amplitude crosses zero.
As a result of the performed calculations it was shown that the minimums of cross sections in scale 
of the atomic mass number $A$ are in the
close neighbourhood with the breaking points of $\alpha_{A}$\footnote{Inverse statement is not true,
the corresponding minimum in cross secion does not exist for every breaking point}
With the increase of the photon virtuality $Q^{2}$,
the location of the minimums moves to the location of breaking points, which means, that they have 
a common nature.\\
\hspace*{1em}
Within the Glauber multiple scattering theory~\cite{A2} the coherent
forward amplitude for the process of diffractive
electroproduction of the vector meson on nucleus
can be written~\cite{A3} as:
\begin{eqnarray}
{f_{A} = 2\pi A f_{N}\int^{\infty}_{0}bdb\int^{\infty}_{-\infty}dz
exp\Big\{- A\frac{\sigma^{VN}_{tot}}{2}
\int^{\infty}_{z}\rho(b,y)dy\Big\}\rho(b,z)e^{i q_{L} z}} \hspace{0.2cm},
\label{eq:Eq1}
\end{eqnarray}
where $f_{N}=i\Im m{f_N}(1-i\alpha_N)$ ($\alpha_N=\Re e{f_N}/\Im m{f_N}$) is the forward amplitude for vector meson
electroproduction by an individual nucleon, $b$ and $z$ are the
impact parameter and longitudinal coordinate of the point in nucleus
where vector meson production took place,
$\sigma^{VN}_{tot}$ is the vector meson - nucleon total cross section.
Longitudinal momentum transfer in the individual
collision at forward direction inside the nucleus is equal
\begin{eqnarray}
{q_{L} = \frac{Q^{2} + M^{2}_{V}}{2\nu}} \hspace{0.2cm},
\label{eq:Eq2}
\end{eqnarray}  
where $\nu$ and $q^{2} = - Q^{2}$ are the energy 
and the square of the four momentum of virtual photon.
In the literature there are usually discussions about the two extreme cases of this variable, so 
called low and high energy limits, i.e. $q_{L} \rightarrow \infty$ and $q_{L} \rightarrow
0$ (see, for instance, Ref.~\cite{A4}). We consider intermediate case of different from zero
and infinity values of $q_{L}$. As an example the results of
calculations performed at values of variables
$Q^2$ and $\nu$ equal to $4 GeV^2$ and $10 GeV$, are presented. According to the
Eq.~\ref{eq:Eq2}, it corresponds to $q_{L} \approx 0.23 GeV$. Its inverse value $l_c$, which is called
the $coherence$ $length$ is equal to $l_c \approx 0.87 fm$.
We show that in this region there is nontrivial dependence of the coherent
differential cross sections from $A$, because of the typical values of $q_{L}r_{A} \sim 3 \div 10$
($r_{A}$ - nuclear radius), which means that nuclear amplitude strongly oscillates (see 
Eq.~\ref{eq:Eq1}).
The forward differential cross section on
nucleus $\Big(\frac{d\sigma}{d\Omega}\Big)_{A}$ is
obtained by taking the absolute magnitude squared of the
amplitude given in Eq.~\ref{eq:Eq1}.
\begin{eqnarray}
\frac{d\sigma^{A}}{d\Omega} = \Big|f_{A}\Big|^{2} \hspace{0.2cm}.
\label{eq:Eq3}
\end{eqnarray}
\begin{eqnarray}
\frac{d\sigma^{A}}{d\Omega} = A^{2}\frac{d\sigma^{N}}{d\Omega}\times \hspace{8cm}
\nonumber
\end{eqnarray}
\begin{eqnarray}
\bigg\{\bigg(\int d^{2}b\int^{\infty}_{-\infty}dz
exp\Big\{- A\frac{\sigma^{VN}_{tot}}{2}
\int^{\infty}_{z}\rho(b,y)dy\Big\}\rho(b,z) cos(q_{L}z)\bigg)^{2} +
\nonumber
\end{eqnarray}
\begin{eqnarray}
\bigg(\int d^{2}b\int^{\infty}_{-\infty}dz
exp\Big\{- A\frac{\sigma^{VN}_{tot}}{2}
\int^{\infty}_{z}\rho(b,y)dy\Big\}\rho(b,z) sin(q_{L}z)\bigg)^{2}\bigg\}
\hspace{0.2cm}.
\label{eq:Eq4}
\end{eqnarray}
The ratio of the real to the imaginary parts of forward amplitude for
electroproduction of
vector mesons on the nucleus is
\begin{eqnarray}
\alpha_{A} = \frac{\Re e{f_{A}}}{\Im m{f_{A}}} = \hspace{8cm}
\nonumber
\end{eqnarray}
\begin{eqnarray}
- \frac{\int d^{2}b\int^{\infty}_{-\infty}dz
exp\Big\{- A\frac{\sigma^{VN}_{tot}}{2}
\int^{\infty}_{z}\rho(b,y)dy\Big\}\rho(b,z) sin(q_{L}z)}
{\int d^{2}b\int^{\infty}_{-\infty}dz
exp\Big\{- A\frac{\sigma^{VN}_{tot}}{2}
\int^{\infty}_{z}\rho(b,y)dy\Big\}\rho(b,z) cos(q_{L}z)} \hspace{0.2cm}.
\label{eq:Eq5}
\end{eqnarray}
Calculations are performed for the case of $\rho^0$-mesons production, and for
$\sigma^{VN}_{tot}$ the value of $25mb$ is used.
Also another values for $\sigma^{VN}_{tot}$ were used to simulate the case of $\phi$ meson
production, as well as to simulate possible $Q^2$ dependence of cross section due to color transparency effect.
Calculations are done for
the nuclear density function in the form of the Woods-Saxon distribution: 
\begin{eqnarray}
{\rho(r) = \rho_{0}/(1 + exp((r - r_{A})/a))} \hspace{0.2cm},
\vspace{0.5cm}
\label{eq:Eq6}
\end{eqnarray}
where parameters are $a = 0.545 fm$ and $r_{A} = 1.14A^{1/3} fm$, the values of
$\rho_{0}$ are determined from the normalization condition
$\int d^{3}r \rho(r) = 1$.
To study the $A$ - dependence one needs to include in consideration large range of nuclei, from 
light to heavy. It is clear, that the description of
whole region by one set of nuclear parameters presented above, leads to the
qualitative representation only.
\begin{figure*}[!htb]
\begin{center}
\epsfxsize=8.cm
\epsfxsize=10.cm
\epsfbox{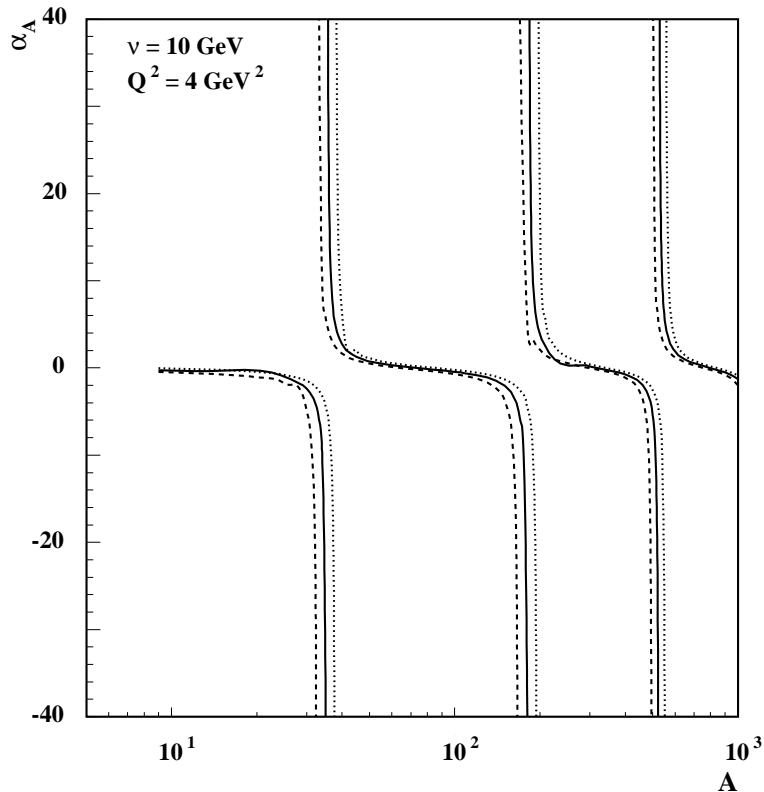}
\end{center}
\caption{\label{xx1}
{\it 
A-dependence of the $\alpha_{A}$, the ratio of the real to the imaginary parts
of nuclear amplitudes, at $\nu = 10 GeV$ and $Q^{2} = 4 GeV^{2}$.
Dashed, solid and dotted curves correspond to the values of $\alpha_{N} = 
- 0.2, 0$ and $0.2$, 
respectively.
 }}
\end{figure*}
On Fig.~\ref{xx1} we present the ratio of the real to the imaginary
parts of the amplitude, $\alpha_{A}$ as a function of $A$ (see Eq.~\ref{eq:Eq5}). 
As was mentioned above, $Q^2$ and $\nu$ were fixed at the
values of $4 GeV^{2}$ and $10 GeV$, respectively
\footnote{For the chosen kinematics the most realistic value of 
$\alpha_{N}$ is
$\alpha_{N} = - 0.2$ (see e.g.~\cite{A5}).}. 
One can see from Fig.~\ref{xx1} that changing of 
$\alpha_{N}$ do not change essentially the shape of curve, but slightly shifts
it from the central position. Taking this into account we will use for the
calculations the value of $\alpha_{N} = 0$ only.
\begin{figure*}[!htb]
\begin{center}
\epsfxsize=8.cm
\epsfxsize=10.cm
\epsfbox{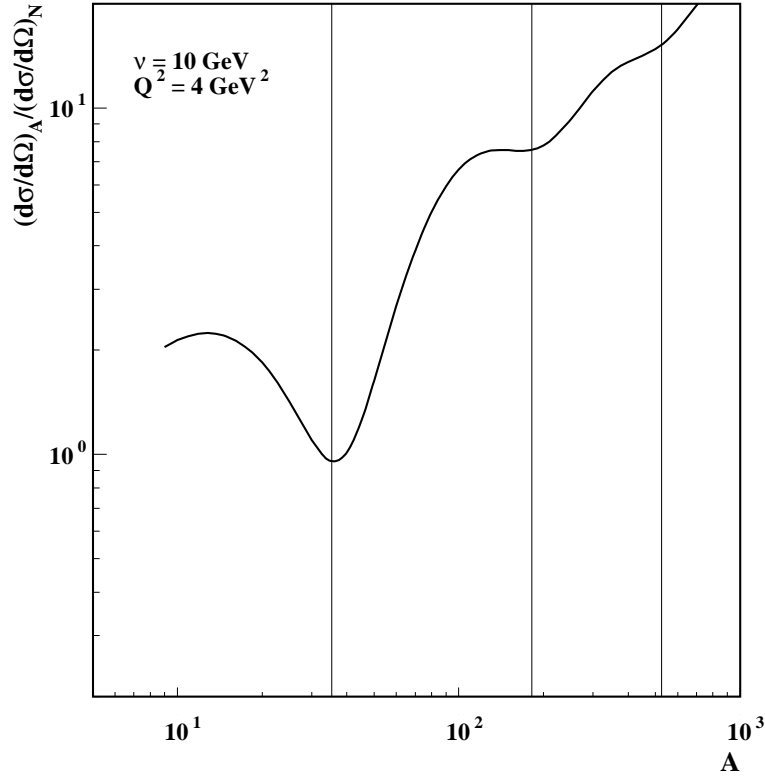}
\end{center}
\caption{\label{xx2}
{\it 
 Ratio of the coherent differential cross sections in forward direction
, as a function of atomic mass number $A$ calculated
at the values of $\nu = 10 GeV$, $Q^{2} = 4 GeV^{2}$ and $\sigma^{VN}_{tot} = 25mb$.
Vertical lines correspond to the $\alpha_{A}$ breaking point positions.
 }}
\end{figure*}
On Fig.~\ref{xx2} the ratio of the coherent differential cross sections in forward direction
, as a function of atomic mass number $A$ is calculated following to Eq.~\ref{eq:Eq4}.
Positions of breaking point coincide with the minimum of cross section, which indicates
on their common nature. Calculations performed with higher values of $Q^{2}$
show that existing minimum moves to the smaller values of $A$ and becomes more
deeper, also, at higher values of $A$, the second minimum arises.
\begin{figure*}[!htb]
\begin{center}
\epsfxsize=8.cm
\epsfxsize=10.cm
\epsfbox{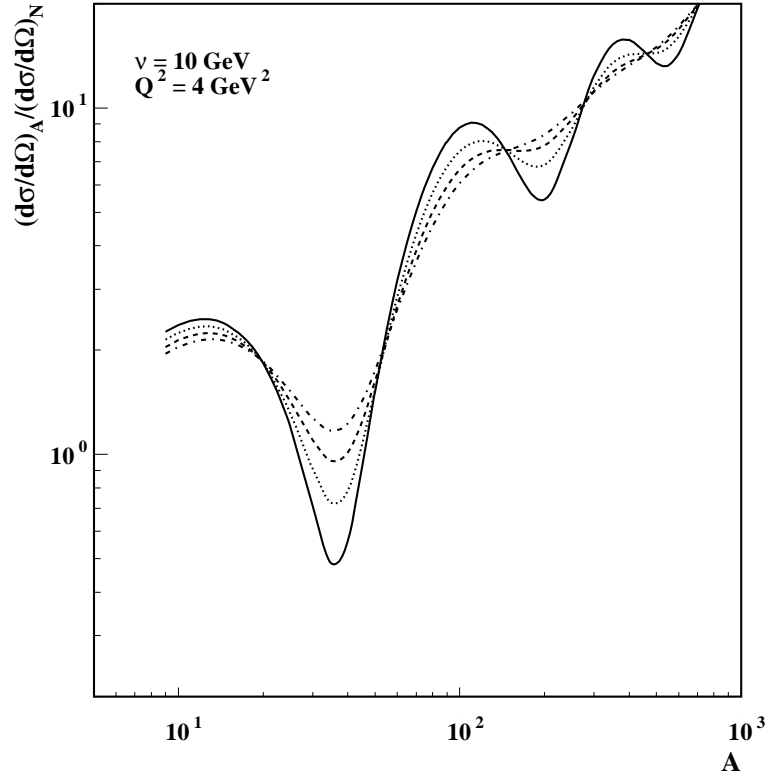}
\end{center}
\caption{\label{xx3}
{\it
 Ratio of the coherent differential cross sections in forward direction
, as a function of atomic mass number A calculated
at the values of $\nu = 10 GeV$ and $Q^{2} = 4 GeV^{2}$. Solid, dotted, dashed and     
dash-dotted curves correspond to the $\sigma^{VN}_{tot}$ values equal to $15, 20, 25$
and $30 mb$, respectively.
 }}
\end{figure*}
On Fig.~\ref{xx3} the
ratio of the coherent differential cross sections in forward direction
$\Big(\frac{d\sigma}{d\Omega}\Big)_{A}\Big/\Big(\frac{d\sigma}{d\Omega}\Big)_{N}$,
as a function of atomic mass number A is calculated
for different values of $\sigma^{VN}_{tot}$. Used values of $\sigma^{VN}_{tot} = 15mb$ and $25mb$ 
are close to the accepted
values for $\phi$ - and $\rho$ - meson - nucleon total cross sections, respectively.\footnote{In order to 
make the strong statement concerning the correspondance to the $\phi$ - meson case we have to change also 
the definition of $q_L$ (see Eq.~\ref{eq:Eq2}), where we use for all cases the $\rho$ -meson mass.} One can 
see, that the nuclear 
cross sections within the local minimums 
are most sensitive to the value of $\sigma^{VN}_{tot}$. It can be used for precise measurement of the
vector meson - nucleon total cross sections. It is interesting to note, that
position of minimum does not depend from the value of cross section.
Although Glauber multiple scattering theory does not take into account the
color transparency effect, some influence ot this effect can be realized by means of variation of the 
average values of $\sigma^{VN}_{tot}$.

Our conclusions are as follows:
\begin{itemize}
\item {
The $A$-dependence of the differential cross sections for the
coherent electroproduction of vector mesons on nuclei in forward direction is studied.
It is shown that they are complicated functions of $A$ with minimums
and maximums.
}
\item {
The ratio of the real to the imaginary parts of the forward amplitude on
nucleus $\alpha_{A} = \Re e{f_{A}} / \Im m{f_{A}}$ has
breaking points at some values of $A$.
}
\item {
Comparison of the $A$ dependence of differential cross sections with $\alpha_{A}$
showes that minimums of cross sections are in the close neighbourhood
with the breaking points of $\alpha_{A}$. With the increasing of photon
virtuality $Q^{2}$,
minimums move to the breaking points.
}
\item {
Nuclear cross sections are very sensitive to the values of
vector meson-nucleon total cross sections and can be used for their
precise determination.
}
\item {One should note that the approach used in this paper can only claim for qualitative
description of the vector mesons coherent
electroproduction in forward direction. We intend to include into consideration in furhter papers also the effects
connected with the various nuclear density distributions; other vector mesons production; as well as
to involve the color transparency effect.}  
\end{itemize}







 

\begin{thebibliography}{99}
\bibitem{A1}  J.G.Asbury et al., Phys.Rev.Lett. {\bf 19} (1967) 865
\bibitem{A2}  R.J.Glauber, In: Lectures in Theor. Phys., v.1, ed.
W.E.Brittin and L.G.Duham. NY: Intersciences, 1959  
\bibitem{A3}  S.D.Drell and J.S.Trefil, Phys.Rev.Lett. {\bf 16} (1966)
 552
\bibitem{A4}  J.Huefner, B.Kopeliovich, J.Nemchik, Phys.Lett. {\bf B383} (1996)362
\bibitem{A5}  T.~Renk, G.~Piller and W.~Weise, Nucl. Phys. {\bf A689} 
(2001), 869
\end{thebibliography}
\end{document}